\documentstyle[12pt,epsfig]{article}
\begin{document}
\title{On the probability distribution of the experimental results}
\author{A.P.Bukhvostov\\
\it  188350, PNPI, Gatchina, Russia }
\maketitle

\begin{abstract}

The analysis of Tables of particle properties shows that the
probability distribution of the results of physical measurements
is far from the conventional Gaussian $\rho(\xi)=exp(-\xi^2/2)
$, but is more likely to follow the simple exponential law
$\rho(\xi)=exp(-\xi)$ ($\xi$ is the deviation of the measured
from the true value in units of the presented standard error).
A gap between the expected and actual probabilities grows with
$\xi$ very rapidly, amounting to $ 10^7 $ at $\xi\approx 6 $,
and is significant even at $\xi=2 $. A more detailed study
reveals the two-component structure of the distribution: the
$exp(-\xi)$ law is closely fulfilled up to $\xi=3$, but then,
at $\xi$ larger than that, the decrease is retarded drastically.
This behaviour can be associated with the existence of two
various types of systematic errors, the detected and undetected
ones. Within some model, both types of errors are seen to affect
the form of the distribution, one at moderate $\xi$ and the
other at large $\xi$. The first type (detected) errors are shown
in some natural-looking assumptions to yield the distribution
not quite equal but close to the simple exponential.
\end{abstract}

\section{Introduction}

It is a usual practice in experimental physics to estimate the probability
for some measured quantity having a certain value with the use of Gaussian
probability distribution. Namely, if the measured value is $A\pm \Delta$,
then the probability that the true value lies in the interval $(a, a+da)$
is commonly found by the formula (Gaussian standard law):
\begin{equation} \label{1}
dw(a)=\frac{da}{\sqrt{2\pi}\,\Delta}\,
e^{\mbox{\large $-\frac{(a-A)^2}{2\Delta^2}$}}.
\end{equation}

The reason for doing so is supposedly provided by the known
central limit theorem asserting the validity of Gaussian law
(\ref{1}) for any random variable satisfying certain rather
general requirements (see e.g. \cite{1}). Surely, the fulfilment
of these requirements as a rule is not analysed in detail,
merely postulated when there are no special reasons for doubt;
but the Gaussian law, whether justified or not, is extensively
used and apparently borne out experimentally in many domains of
statistics.

A characteristic feature of the Gaussian distribution is very rapid
fall-off of the probability with the departure from the most probable
value. Defining the normalized deviation
\begin{equation} \label{2}
\xi=\frac{|a-A|}{\Delta}
\end{equation}
and the integrated probability from (\ref{1})
\begin{equation} \label{3} w(\xi)=
\sqrt{\frac{2}{\pi}}\int_{\xi}^{\infty}dt\ e^{-\textstyle
\frac{t^2}{2}}=1-\Phi\left(\frac{\xi}{\sqrt{2}}\right)
\end{equation}
($\Phi$ is the known probability integral),
we have e.g. $w(4)\sim 10^{-4}$, \mbox{$w(5)\sim 10^{-6}$},
$w(6)\sim 10^{-9}$, etc. These numbers are too small to be taken
seriously, and they have hardly been ever put to a severe
experimental test; one can only argue to have checked the Gaussian
distribution in the region of moderate deviations
($\xi\widetilde{<}4$) where probabilities are measurable. Still
Gaussian estimates are widely used also in that faraway region
($\xi\widetilde{>}6$), evidently for the lack of a better method.

In particle physics, however, the measurement results deviating
from exact values by as much as 6 errors or more are anything
but an extreme rarity, their probability has in fact nothing to do
with such tiny numbers as $10^{-9}$ and amounts to a quantity of
order $10^{-2}$ (as estimated from the total number of data
$\sim 10^{3}$ in Tables of particle properties). This huge
discrepancy (about 7 orders), taken literally, is not easy to
eliminate within the Gaussian-like behaving distribution
functions. A suggestion had been put forward (hardly in
earnest), and gained popularity, that a true estimate of
probability should be obtained with substituting for the
dispersion $\Delta$ in eq.  (\ref{1}) an experimental error
multiplied by 3, since \vspace{-2mm}
$$exp{\left(-\frac{1}{2}\cdot 6^2\right)}\approx
10^{-7}exp{\left(-\frac{1}{2}\cdot\left(\frac{6}{3}\right)^{2}\right)}.$$

\vspace{-2mm}
The scale factor 3 for the error looks unreasonably big, but regardless of
its exact value, this primitive approach appears unsatisfactory because of
overstating the significance of a few wrong or inaccurate measurements
with large $\xi$ in the bulk of much better experimental data. Large
deviations are obviously due to systematic errors which plague many
modern ingeniously designed complex experiments and are difficult
to estimate or even to unveil. As no general method of doing this
is accepted and every experimentalist uses here his own discretion,
there is little reason to believe a priori in Gaussian or some
other form of the distribution in question, it is moreover unclear
even whether such distribution of any kind should exist at all.

In 1973 the author did the work \cite{2} intended to learn the
probability distribution of measurement results with $\xi$ in
experimental way, by closer examination of  Tables  of particle
properties. It was found that the actual distribution (with the
understanding that it does exist) showed the behaviour quite
different from Gaussian, decreasing with $\xi$ much more slowly.
Remarkably, the form of the distribution appeared to agree
fairly well with the simple formula
\begin{equation} \label{4}  w(\xi)=e^{-\,\xi}, \end{equation}
or, in terms of eq.(\ref{1}),
\begin{equation} \label{5}
dw(a)=\frac{da}{\Delta}\ e^{\mbox{\large $-\frac{|a-A|}{\Delta}$}}.
\end{equation}
Noteworthy is the absence of a possible number coefficient in front
of $\xi$ in the exponent, i.e. this coefficient proved to equal 1
with reasonable accuracy.

In a related work by M.Roos et al. \cite{3} the problem was treated
somewhat differently. These authors also detected large departure
of the real data distribution from Gaussian but gave preference to a
power-behaving (Student-like) fit involving two free parameters (see
eq. (\ref{13}) of Sec.~2). While looking less attractive than our
eq. (\ref{4}), it provided better conformity with the actual
distribution; however, the arguments to its favour are not convincing
and do not allow one to decide between this and some other fit
of a similar kind. The fact is that the form of the fitting
function is not specified by a theory and can be chosen more or
less arbitrarily. The present work in its major part is aimed at
finding this functional form from some probable-looking model
assumptions.

The attempts to explain the non-Gaussian behaviour of data distribution
made in both works \cite{2,3}, though different in resulting formulae,
were conceptually similar and based on the idea of taking into account
the inevitable uncertainty in the experimentally determined errors.
The ambiguity of the result is a consequence of essential dependence
on the assumed form of the unknown probability distribution for the
errors. Here is the weak point of this approach: the effect of the
error uncertainty, surely existing and significant as it is, escapes
solid quantitative estimates and can provide only a crude explanation
of the Gaussian law violation, defying accurate experimental tests.

It will be seen however that there exists another, in a sense competitive
effect which is more liable to evaluation and may appear dominant
at some~$\xi$. It can be designated the variable dispersion
effect and is operative even with fixed and exactly known mean squared
error, provided the dispersion of the measuring device is not constant
but depends on the value of the measured quantity. While the first
effect (the error uncertainty) is caused mainly by the systematic
errors overlooked or radically underestimated by the experimentalists,
the second effect can be attributed to the influence of those
systematic errors which are detected and taken properly into account,
but nevertheless induce a distortion of Gaussian law in default of
compliance with the conditions of central limit theorem.

In this work the experimental data probability distribution is
reexamined on the basis of a later and more abundant statistical
material. It turns out that the distribution exhibits a well-marked
two-component structure characteristic of the superposition of two
functions with different behaviour (see fig. 2, curve 1 to be discussed
below): one fall-off regime is changed rather abruptly by the other
near the point $\xi=3$. It seems natural to associate these two regimes
with two various mechanisms whose contributions have different $\xi$
dependence and become comparable at $\xi\approx 3$. The right-hand
portion of the distribution curve ($\xi>3$) which was not quite clearly
visible with the earlier poor statistics may well have been produced
by the effect of error uncertainty, while the left-hand portion
following closely the simple-exponential law of eq. (\ref{4}) could
be shaped with the variable dispersion effect which will be seen to
yield a distribution of a similar form. Of particular interest is
the fact that the unit coefficient in the exponent gets a natural
explanation in this scheme.

The paper is organized as follows. Sec.~2 contains an account of
the method and results of experimental determination of the data
probability distribution with the use of Tables of particle properties.
The remaining part of the paper is concerned with the
theoretical interpretation of the obtained distribution. In
Sec.~3 the effect of the error uncertainty is considered. Sec.~4
is devoted to a cursory discussion of the central limit theorem
and a possible way of its expansion. In Sec.~5 which is the main
in the work, a model of the variable dispersion effect is
developed and analysed.  It is shown that under some rather
general assumptions the asymptotic form of the distribution at
$\xi\gg 1$ must be $\xi^{-2} exp{(-\xi)}$. The last Sec.~6
contains a short review of the results and some concluding
remarks.

\section{Experimental distribution}

For experimental evaluation of the data probability distribution
all which is needed is a sufficiently abundant set of
measurements of some quantity with the exactly known true value.
Of course, the true value is normally unknown during the
measurement, but we can restrict our consideration to the old
experiments for which the later, much more accurate results are
now available to be taken for the exact values. As for the
number of measurements, although it is hardly ever large enough
when dealing with only one definite physical quantity, it can well
be made statistically significant by using the hypothesis of
universality of the distribution, i.e. assuming the probability
$w(\xi)$ to depend only on $\xi$ and nothing else. The
probability density will then agree with the formulae
\begin{eqnarray}
dw(a)=\frac{da}{\Delta} \rho(\xi),\\ \label{6}
\rho(\xi)=-w\,^\prime (\xi), \label{7}
\end{eqnarray}
generalizing eqs. (\ref{1}) and (\ref{3}). This hypothesis, if true
(which point will be returned to below), allows one to take into account
all measurements with the same $\xi$ on equal terms, lumping together
the results of quite dissimilar experiments, irrespective of the quantity
measured or the technique employed. Note that we suppose the
distribution to be symmetric about its centre, since $\xi$ defined
by eq. (\ref{2}) is an even function of $(a-A)$, so we should avoid
consideration of measurements with manifestly asymmetric errors.

The distribution function $w(\xi)$ designating the probability for the
measurement result to deviate from the true quantity by $\xi$ or more
standard errors, can thus be found from any set of data by the formula
\begin{equation} \label{8}
w(\xi)=\frac{1}{N}
\sum_{i=1}^{N}\theta\left(\frac{|A_i-a_i|}{\Delta_i}-\xi\right),
\end{equation}
where $N$ is the total number of data in the set (supposed to be
very large), $a_i$ is the exact value of the quantity measured
in $i$-th experiment, $A_i$ and $\Delta_i$ are its measured
value and the error, and $\theta(x)$ denotes the Heavyside step
function (1 for $x>0$ and 0 for $x<0$). When using Tables of
particle properties as a data source, $A_i$ and $\Delta_i$ are
suggested to be taken from an earlier (`old') Table, $a_i$ from
the latest (`new') Table. So in practice the values of $a_i$ are
themselves not quite exact but approximate with the new errors
$\delta_i$, and can be treated as nearly exact only when a new
error $\delta_i$ is small compared to the corresponding old error
$\Delta_i$, which is not always the case. The influence of
uncertainty of $a_i$ can be suppressed by including in the set
only the data with sufficiently large ratio $\Delta_i/\delta_i$;
in what follows a rather liberal data cutoff is used
\begin{equation} \label{9}
\frac{\Delta_i}{\delta_i}\ge 2.5\,. \end{equation}
(In \cite{2} changing the cutoff from 2.5 to 4 was found to
reduce the number of data $N$ by about 10\% with quite
negligible effect on the distribution).

The described method of finding the function $w(\xi)$ was
applied first in 1973~\cite{2} to the Tables of 1964~\cite{4}
and now once again to the Tables of 1978~\cite{6} (the old
data), with `exact' values extracted from the new Tables of
1972~\cite{5} and 1994~\cite{7} respectively. Apart from the
restriction (\ref{9}), the procedure of data selection included
yet some more criteria. The following data types were rejected:

---  the results treated by the compilers as unreliable or
preliminary (enclosed in parentheses);

---  the data taken from review articles (marked by the label
RVUE);

---  the results of measurements of essentially positive (by
physical meaning) quantities when the error exceeded half the
measured value;

---  the data with asymmetric errors of the form
$A^{+\Delta_1}_{-\Delta_2}$ when $\Delta_1$ and $\Delta_2$
differed by more than 10\% (otherwise the result was accepted
with the error taken to be symmetric and equal to the greater of
$\Delta_1$ and $\Delta_2$);

---  all data relating to wide resonances, $\Gamma>120$ MeV.
In \cite{2}, moreover, no resonance data at all were considered,
only those relevant to strong-interaction stable particles.

We are not dwelling on the discussion of these constraints, aimed
mainly at lowering the contribution of dubious or too uncertain results
and of some temporal changes in general concepts (e.g. in the common
definition of the mass and width of very wide resonances). One may argue
if the variation (strengthening or loosening) of formulated criteria
should really have no effect but the change of the total normalization;
we shall touch on this point once more in the next section.

It must be emphasized that the total statistical sample of 933 data
examined in this work includes all measurement results contained in
the Table \cite{5}, irrespective of the quantity taken, except only
for those rejected by the above-listed restrictions. Its inspection
shows that it comprises four groups of measurements having comparable
weights (classified by the quantity that was measured): measurements
of particle masses (223 data, or 24\%), lifetimes or widths (196 data,
or 21\%), branching ratios of various decay modes (316 data, or 34\%),
and the last group (198 data, or 21\%) combines all other particle
characteristics (magnetic moments, form factors, parameters of angular
and energy distributions of decay products) as well as some interaction
constants ($g_A/g_V$, parameters of $CP$-nonconservation etc.). No
account was taken of possible correlation between various quantities
obtained from one experiment, they were taken to be independent if the
compilers used them in statistical averaging. The number of such
many-data experiments does not seem to be very small, this effect may
cause some distortion and should be kept in mind. Another analogous
interfering factor could be the mutual influence and correlation
between the results of various experimental groups which are ideally
viewed as quite independent. The significance of these effects is
difficult to consider quantitatively, but one may hope it would be
true to ignore them in the first approximation.

The results of the described treatment of Particle Data Tables are
shown in figs.~1 and~2 (the curves labelled 1). Fig.~1
reproduces the earlier finding~\cite{2} and is given here for
convenience of comparison; the newly obtained distribution of
data from~\cite{5} is presented in fig.~2. The difference is seen
to be modest and can well be assigned to damping of statistical
fluctuations with growth of the total statistics from 209 to 933;
however, it should be noted that there is a certain systematic
excess of the distribution in fig.~2 over that in fig.~1 which
may appear to be evidence against the above-formulated
universality. Here we shall not discuss this point in much
detail, but turn now to the other, much more pronounced features
of the resulting distribution.

First of all, the real distribution is apparent to differ
radically from Gaussian (curve 2 in figs. 1 and 2), especially
at large $\xi$, the fact already noted above. The difference
grows quickly with $\xi$ and ranges up to many orders at
$\xi\widetilde{>}6$, but is rather large even for moderate
values of $\xi$ ($\sim 3$~times at $\xi=2$, $\sim 20$~times at
$\xi=3$). It is clear that Gaussian estimates highly underrate
the deviation probabilities nearly everywhere in $\xi$ and are
entirely misleading.

An essential fact here is that the real distribution curve 1
not only passes much higher than the Gaussian curve 2, but
moreover exhibits a distinctly different behaviour: the two
curves have the opposite convexity (in log scale), i.e. the real
probability decay rate $d|ln\,w|/d\xi$ decreases with $\xi$,
while by Gaussian law it should be growing $\sim\xi$. It follows
that there is little point in modifying Gaussian law (\ref{1}) with
only renormalizing the error $\Delta$ by any constant factor, as
this procedure leads to nothing more than stretching the curve along
the $\xi$ axis with no effect on convexity. The curves 4 and 5 in
fig. 1 show that doubling, the more so tripling the error (i.e.
replacing $\xi$ in eq. (\ref{3}) by $\xi/2$ or $\xi/3$) results in
grossly overestimating the deviation probabilities against their
experimental values in the region of moderate $\xi$. Clearly, the
agreement cannot be achieved by any choice of the error
renormalization factor and the Gaussian-like approach itself
does not work.

It is natural to try for the experimental distribution the
linear-exponential approximation (\ref{4}) (straight line 3 in
figs.~1 and~2) which appears to fit rather well, better than one
could expect without any prerequisite. Specifically, a fact
deserving consideration is the unit slope of the fitting curve,
which corresponds to the above-mentioned unit coefficient in the
exponent of eq. (\ref{4}). Of course, it may well be a mere
coincidence, but it is suggestive to see here a regular phenomenon
and to search for its origin.
\begin{figure}
\centerline{\epsfig{file=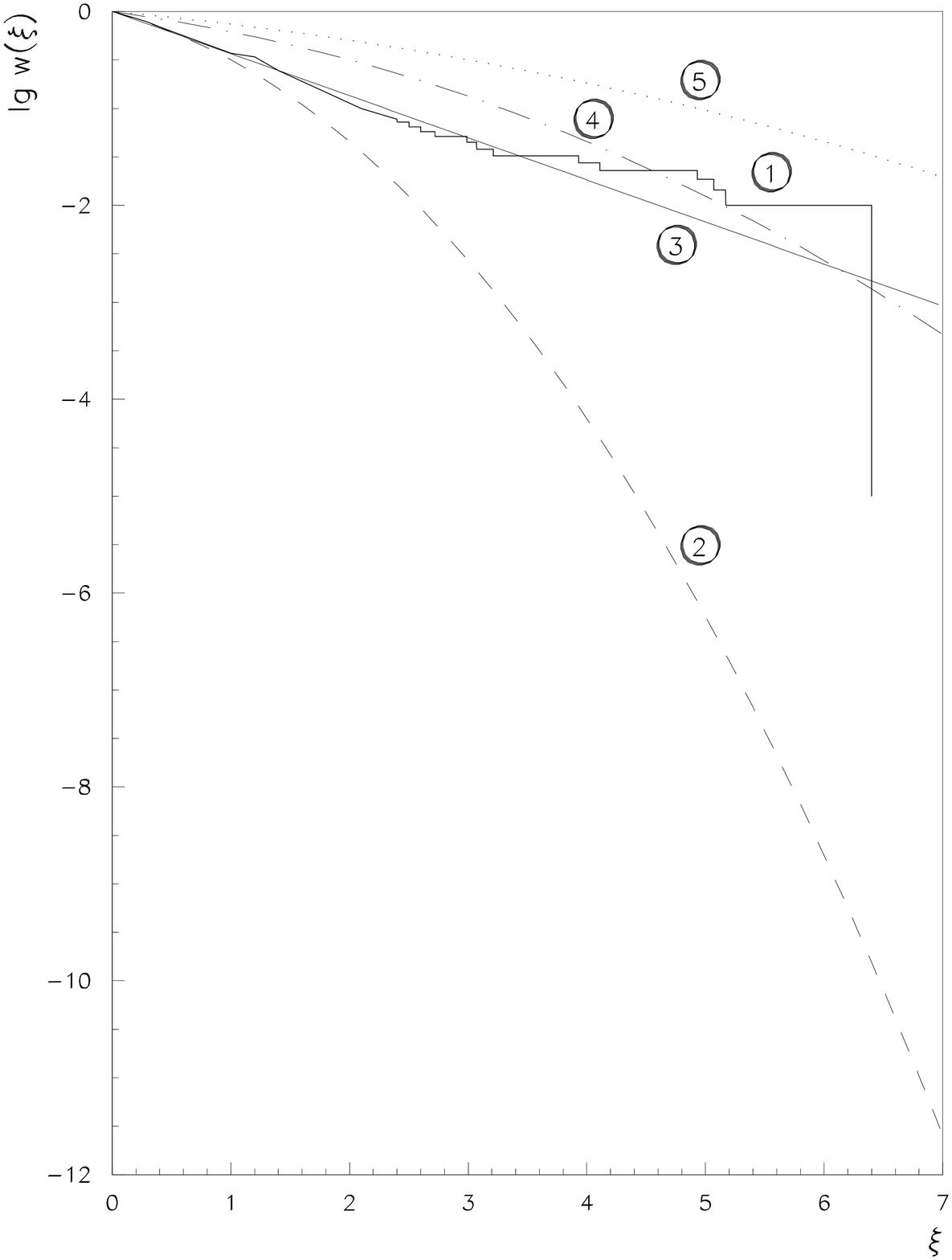,width=16cm}}
\begin{caption}\,$\bigcirc\!\!\!\!1$
---experimental distribution of 209 measurement results,
ref.~\cite{2}; $\bigcirc\!\!\!\!2$ ---Gaussian standard law,
eq.~(\ref{3}); $\bigcirc\!\!\!\!3$ ---simple exponential law,
eq.~(\ref{4}); $\bigcirc\!\!\!\!4$\,,$\bigcirc\!\!\!\!5$
---Gaussian distribution with doubled and tripled dispersion.
\end{caption} \end{figure}

Closer inspection of fig. 2 shows that the linear-exponential fit
(\ref{4}) being very good at $\xi<3$ breaks down at large $\xi$
where the real probability curve~1 displays essentially slower
fall-off. It seems that the point $\xi=3$ marks the region where
the fall-off regime undergoes a change closely resembling that of
the decay curve of a two-component radioactive substance. A limited
statistics (65 measurements with $\xi>3$ of the total 933) does
not allow one to recognize the functional form of the decrease
at large $\xi$, but its slowing down is seen clearly enough, in
distinction to fig.~1 where the statistics was too meagre for
that (9 measurements with $\xi>3$ of the total 209). As a simple
illustrative example, the function with a similar behaviour
\begin{equation} \label{10}
w_1(\xi)=\frac{e^{-\,\xi}+C}{1+C}, \;\;\;\;\;C=0.05
\end{equation}
is displayed in fig. 2 (curve 4). The value of the constant $C$ is
chosen so that the two terms in the numerator were equal at
the breakpoint $\xi=3$,
the denominator serves to ensure the normalization condition
\begin{equation} \label{11}
w_1(0)=\int_0^\infty \rho(\xi)d\xi=1.
\end{equation}
Of course, this function cannot represent the real distribution
at all $\xi$ since it does not vanish at $\xi\to\infty$, but can be
considered as a reasonable approximation for moderate $\xi$ when
the `long-lived' component decays very slowly and its change is
inappreciable in the range of $\xi$ under study.

It should be noted that both approximations (\ref{4}) and
(\ref{10}) are expected to fail also at small $\xi$, because
they do not satisfy the condition
\begin{equation} \label{12}
\rho^\prime (0)=-\left.\frac{d^2 w}{d\xi^2}\right|_{\xi=0}=0
\end{equation}
which must be fulfilled if the probability density $\rho(\xi)$
is even in $\xi$ and analytic at the point $\xi=0$. In fact at
small $\xi$ the distribution follows Gaussian rather than
linear-exponential law, which can be ascertained by viewing
fig.~2 at some magnification; the difference is slight but
unambiguous in favour of Gaussian behaviour.
\begin{figure}
\centerline{\epsfig{file=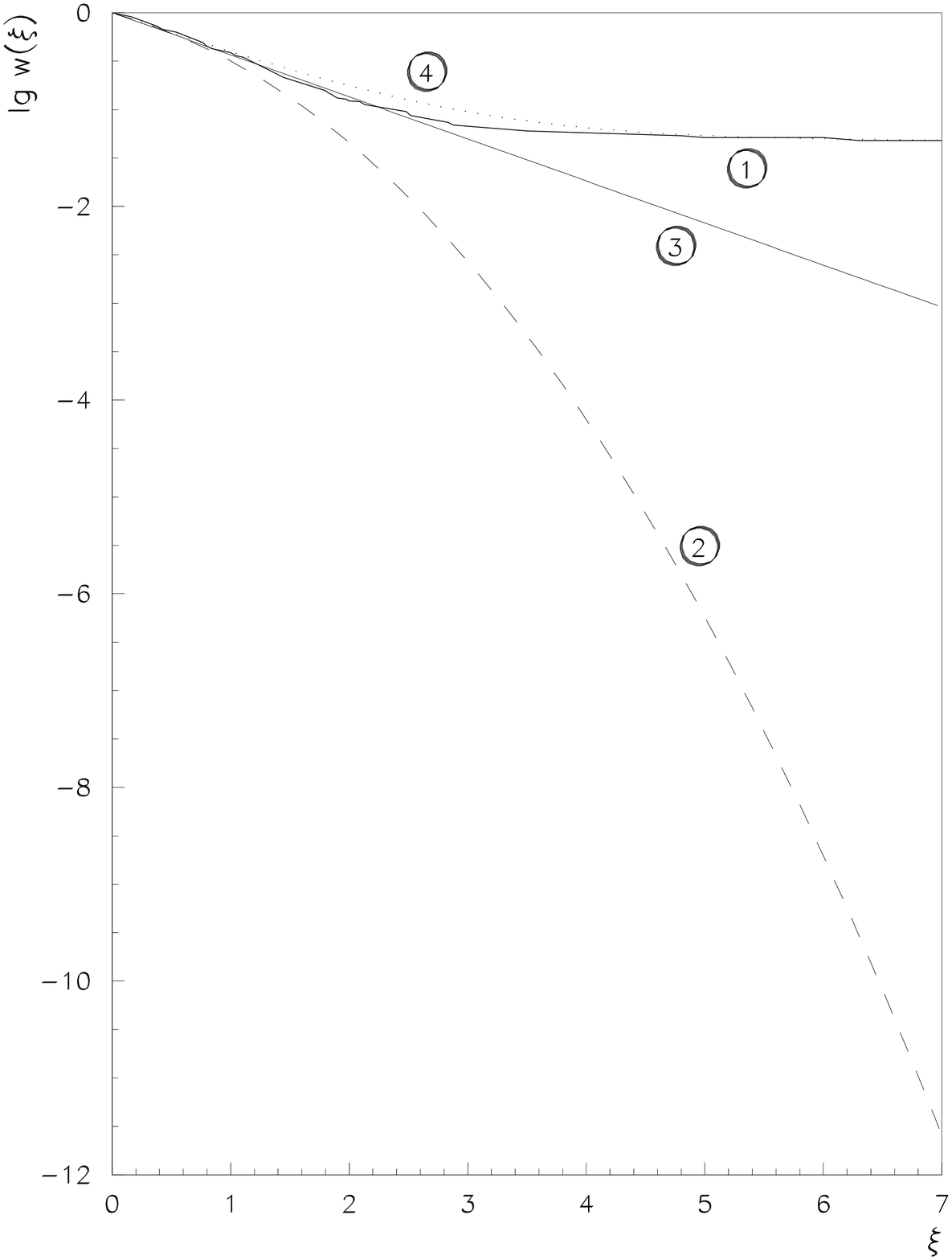,width=16cm}}
\begin{caption}\,$\bigcirc\!\!\!\!1$
---new experimental distribution of 993 measurement results from
ref.~\cite{7}; $\bigcirc\!\!\!\!2$\,,$\bigcirc\!\!\!\!3$ ---same as in
fig.~1; $\bigcirc\!\!\!\!4$ --- modified exponential distribution,
eq.~(\ref{10}).\end{caption}\end{figure}

Before proceeding to the interpretation of the indicated
features of the experimental data probability distribution, it
will be pertinent to touch briefly on the work \cite{3}
concerned with similar problems. Its authors had a well-defined
practical aim---to develop a sound procedure for averaging
together the poorly reconcilable measurement results obtained by
various experimental groups, and doing this required information
on the real data distribution. To get it, they used a procedure
differing from ours in some respects. First, they made efforts
to possibly lower the influence of systematic errors and tried
to select only those measurements which would not be subject to
them. Second, they had to do with only one table (the latest one)
and no later tables, so in place of the true value $a$ in
eq.~(\ref{2}) they used a weighted mean calculated by averaging
the related measurement results from the same table. Thus their
distribution does not quite coincide with ours and some
difference would appear natural. It is worth noting that they
did not merely postulate the universality condition (\ref{6})
but put it to some check, separating the total sample of 306
measurements into several groups and comparing the distributions
in various groups; all of them proved to be essentially the same
within the probable statistical fluctuations.

In fig. 3 curve 1 shows the total data distribution from~\cite{3}
represented in the form suitable for comparison with figs.~1
and~2 (in~\cite{3} a histogram for the probability density
$\rho(\xi)$ was given; here it is recast into the integral
distribution $w(\xi)$ with some interpolation). It can be
observed to bear certain similarities to the distribution in
fig.~2: a remarkable excess over the Gaussian distribution
(curve 2), an approximately linear-exponential decrease at
intermediate $\xi$, and bendings (decay slowing down) both at
small and large $\xi$ (the latter being not quite clearly seen
probably owing to deficient statistics, only 1.5 times that in
fig.~1). There are also clear-cut differences: the large $\xi$
bending (if any) occurs later in $\xi$, and in the linear-exponential
region the slope is about 1.3 instead of unity (i.e. $w(\xi)\sim
exp\,(-1.3\,\xi) \,)$, so approximation (\ref{4}) (curve 3) is
here not good. On the whole, the distribution looks as a kind of
interpolation between the Gaussian law and something like eq.~
(\ref{10}) (curves 2 and 4 in fig.~2).
\begin{figure}
\centerline{\epsfig{file=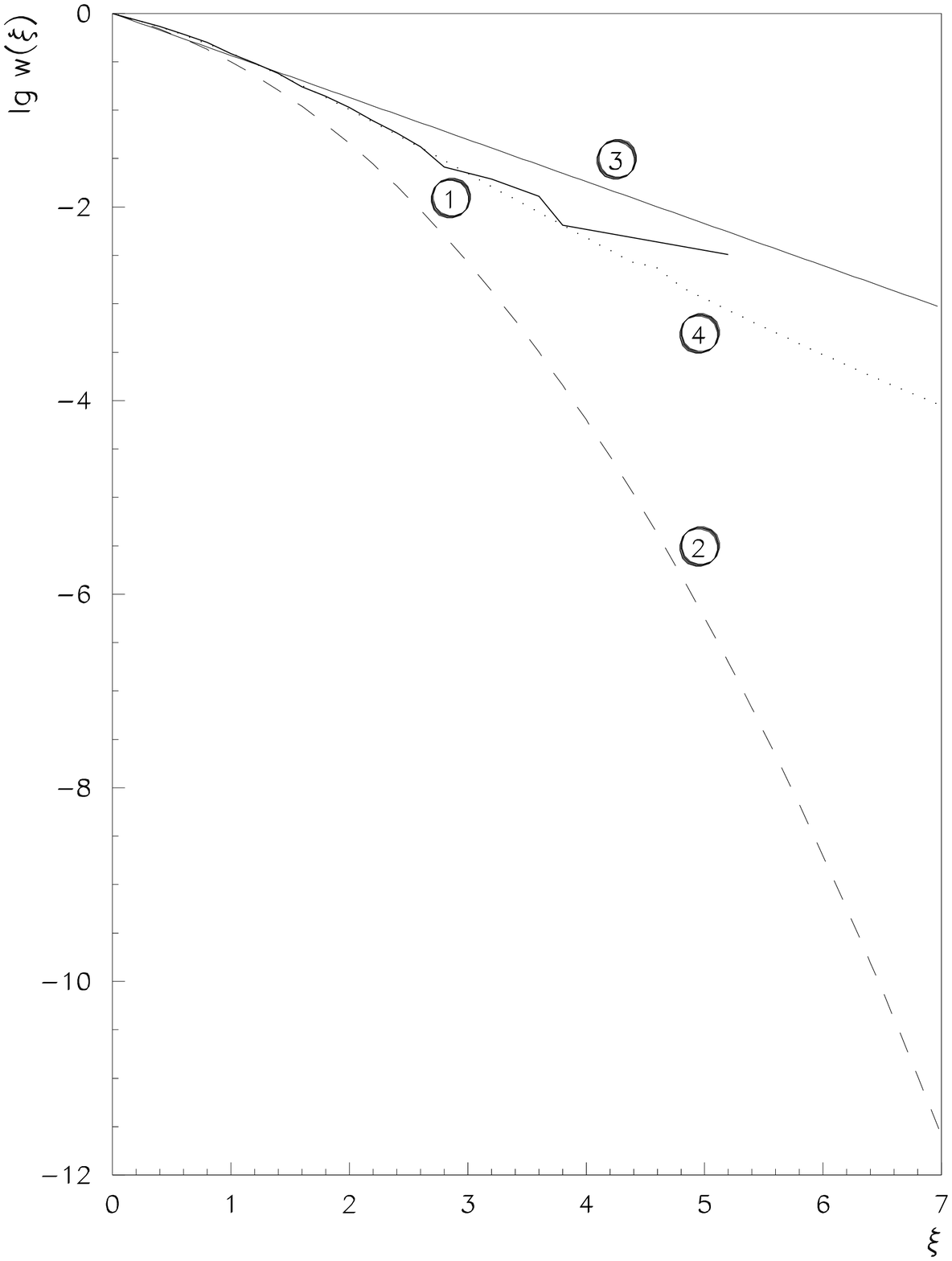,width=16cm}}
\begin{caption}\,$\bigcirc\!\!\!\!1$
---distribution of 306 measurement results from
ref.~\cite{3}; $\bigcirc\!\!\!\!2$\,,$\bigcirc\!\!\!\!3$ ---same as in
fig.~1; $\bigcirc\!\!\!\!4$ --- modified Student's distribution,
eq.~(\ref{13}).\end{caption}\end{figure}

For a fitting function to describe the obtained distribution,
there was taken in~\cite{3} a slight modification of Student's
formula \begin{equation} \label{13} \rho(\xi)=
K\left(1+\frac{\xi^2}{nc^2}\right)^{-\textstyle\frac{n+1}{2}}
\end{equation}
containing two free parameters, $n$ and $c$ (normalization constant
$K$ is defined from eq.~(\ref{11})). With $c=1$ and integer n,
eq.~(\ref{13}) represents the probability density of a random quantity
\begin{equation} \label{14} \xi=\sqrt{\frac{n}{n+1}}\;
\frac{\displaystyle\sum_{i=1}^{n+1} x_i}
{ \displaystyle\sqrt{\sum_{i=1}^{n+1}
\left( x_i-\frac{1}{n+1}{\sum_{k=1}^{n+1} x_k}\right) ^2}}\;,
\end{equation}
which is a function of $(n+1)$ random variables $x_i$ all having
identical normal distributions with zero mean value (or alternatively
they can be viewed as the results of $(n+1)$ independent measurements
of a random variable normally distributed about zero). A scaling
parameter $c$ is introduced to take into account a possible
underestimate of the measurement error by the experimentalists
just in the manner discussed in Introduction. The values of the
parameters adequate to experimental distribution were found to be
\vspace{-2mm} $$n=10,\;\; c=1.11 \,.$$
\vspace{-6mm}

The corresponding integral distribution $w(\xi)$ is represented by
curve~4 in fig.~3 which practically coincides with the experimental
curve~1 nearly everywhere (except for large $\xi$ where the
statistics is scanty). Notwithstanding this coincidence, the form
of eq.~(\ref{13}) does not look convincing; below the arguments
will be adduced for another dependence, more closely resembling
(but not identical to) eq.~(\ref{10}).

\section{The effect of the error uncertainty}

The most prominent feature of all distributions considered above
(figs.~1--3) is their large excess over the Gaussian law predictions,
and its explanation is natural to try to begin with taking into account
the obvious fact that the experimental error is never measured exactly
but may differ significantly from the true dispersion of the measuring
device. Such approach was applied in~\cite{2} and in a less explicit
form in~\cite{3}; here it will be reexamined in some detail.

Let us consider a model based on the following postulates:

1. Each measurement device or experimental setup is subject to
fluctuations producing at the output in place of the measured quantity
$a$ a random quantity $x$ spread with some probability density $p(x,a)$
which is an individual characteristic of a given device. In other
words, when measuring a quantity $a$, one obtains the result confined
between $x$ and $x+dx$ with the probability $p(x,a)\,dx$. The
probability density $p(x,a)$ is assumed to have a Gaussian form
\begin{equation} \label{15}
p(x,a)=\frac{1}{\sqrt{2\pi}\,\sigma}\,
 e^{\mbox{\large $-\frac{(x-a-s)^2}{2\sigma^2}$}}
\end{equation}
with two parameters $s$ (scale shift) and $\sigma$ (dispersion)
characterizing a given device.

2. There exists in the world a huge number $M$ of devices distributed
over the parameters $s$ and $\sigma$ with the density $P(s,\sigma\!)$, so
that the number of devices with the values of the parameters between
$(s,\sigma\!)$ and $(s+ds,\sigma+d\sigma\!)$ is
$dM=MP(s,\sigma\!)\,ds\,d\sigma$.

3. The parameters $s$ and $\sigma$ of any given device are not known
exactly but measured with a limited accuracy and found, instead of
$(s,\sigma\!)$, to be in the range of $(s',\sigma')$ to
$(s'+ds',\sigma'+d\sigma')$ with the probability
$q(s,\sigma|s',\sigma')ds'd\sigma'$. The probability density
$q(s,\sigma|s',\sigma')$ is supposed to be i) a function of
the difference $(s-s')$, not of $s$ and $s'$ separately, and ii) a
homogeneous function of its four arguments, i.e.
\begin{equation} \label{16}
q(s,\sigma|s',\sigma')=\frac{1}{\sigma^2}\:
q\!\left(\frac{s'-s}{\sigma},\,\frac{\sigma'}{\sigma}\right),
\end{equation}
where $q(u,t)$ is dimensionless.

In such model the experimental distribution of measurement results with
the quantity
\begin{equation} \label{17}
\xi=\frac{x-s'-a}{\sigma'}\end{equation}
will be
\begin{eqnarray} \label{18}
\frac{dw(\xi)}{d\xi}\!&=&\!\!\int\!P(s,\sigma\!)ds\,d\sigma\!\int\!
q(s,\sigma|s',\sigma')ds'd\sigma'\!\int\!dx\,p(x,a)\,
\delta\!\!\left(\xi-\frac{x-s'-a}{\sigma'}\right)\nonumber\\
&=&\int\!P(s,\sigma\!)ds\,d\sigma\!\int\!q(s,\sigma|s',\sigma')ds'd\sigma'
\frac{\sigma'}{\sqrt{2\pi}\sigma}\,e^{\mbox{\large
$-\frac{(\xi\sigma'+s'-s)^2}{2\sigma^2}$}}.
\end{eqnarray}
Note that $\xi$ as defined by eq.~(\ref{17}) slightly differs from that
of eq.~(\ref{2}) and is no more a positive definite quantity; to come
back to former definition in eq.~(\ref{15}) and analogous formulae
below a symmetrization with respect to $\xi\to-\,\xi$ should be made.

Using eq.~(\ref{17}) and introducing dimensionless variables
\begin{equation} \label{19}
u=\frac{s'-s}{\sigma}\;,\;\;\;t=\frac{\sigma'}{\sigma}\;,
\end{equation}
we can separate the integral over $ds\,d\sigma$ and take it in view of
the normalization condition
\begin{equation} \label{20}
\int\!P(s,\sigma\!)ds\,d\sigma=1,\end{equation}
so, as might be expected, the distribution of devices $P(s,\sigma\!)$
drops out of the formula and eq.~(\ref{18}) becomes
\begin{equation} \label{21}
\rho(\xi)=\frac{1}{\sqrt{2\pi}}\int du\,dt\,q(u,t)\,
te^{\mbox{\small $-\frac{1}{2}(u+t\,\xi)^2$}}.
\end{equation}

An essential input on the right of eq.~(\ref{21}) is the function
$q(u,t)$ governing the accuracy of determination of the measurement
error $\sigma'$, the resulting distribution being highly dependent on
the suggested properties of this unknown function. Clearly, taking
\begin{equation} \label{22}
q(u,t)=\delta(u)\,\delta(t-1), \end{equation}
we arrive at Gaussian distribution (\ref{3}). Replacing $\delta(t-1)$
by $\delta(t-k)$ (with $k$ constant) gives the scaled Gaussian law
discussed in Introduction. Any kind of smearing the $\delta$-function
yields generally some departure from normal distribution, and there is
nothing to prevent from getting any reasonable expression for the
result by choosing an appropriate form of $q(u,t)$. In particular,
putting \begin{equation} \label{23}
q(u,t)=\delta(u)\,\frac{\mbox{\small$\alpha$}}{t^3}\,
e^{\mbox{\large$-\frac{\alpha}{2t^2}$}},\end{equation}
we get from eqs. (\ref{21}) and (\ref{7})
\begin{equation} \label{24}
w(\xi)=\frac{1}{\sqrt{\mbox{\small$\alpha$}}}\,e^{-\,\xi\,\sqrt{\alpha}},
\end{equation}
which at $\alpha=1$ is just eq.(\ref{4}). It can be remarked that if
$q(u,t)$ has the form $\delta(u)q(t)$ with $q(t)$ smooth and vanishing
both at $t=0$ and $t\to\infty$, then the small $\xi$ behaviour of
$w(\xi)$ is determined in general by the large $\xi$ asymptotics of
$q(t)$, and conversely, the asymptotics of $w(\xi)$ at $\xi\to\infty$
depends on the behaviour of $q(t)$ at $t=0$.

Among other possible expressions for the function $q(u,t)$ of interest
is that one which corresponds to the parameters $c$ and $\sigma$ of a
device being determined from a certain number $n$ of independent
measurements of some known quantity~$a$. If the results of these
calibrating measurements are $x_i\;\; (i=1,...\,n)$, then
\begin{equation} \label{25}
s'=\frac{1}{n}\,\sum_{i=1}^{n}x_i-a\,,\;\;\;
\sigma'\,^2=\frac{1}{n-1}\sum_{i=1}^{n}(x_i-a-s')^2, \end{equation}
so, in view of the relation $d\sigma'\,^2=2\sigma'd\sigma'$ and
eq.~(\ref{15}),
\begin{eqnarray} \label{26}
q(s,\sigma|s',\sigma')=2\sigma'\int\frac{d^n x}{(\sqrt{2\pi}\sigma\!)^n}\,
e^{\mbox{\large$-\frac{1}{2\sigma^2}\scriptstyle\sum_i$}
\mbox{\small$(x_i-a-s)^2$}}\cdot \nonumber\\
\textstyle\delta (s'+a\mbox{\large$-\frac{1}{n}$}\sum_i {x_i})\,
\delta(\sigma'\,^2\mbox{\large$-\frac{1}{n-1}$}\sum_i{(x_i-a-s')^2})\\
\nonumber\\
\sim\frac{\sigma'\,^{n-2}}{\sigma^n}\,
e^{\mbox{\large$-\frac{1}{2\sigma^2}$}
{\mbox{\small$[(n-1)\sigma^2+n(s'-s)^2]$}}},\,\nonumber
\end{eqnarray}
where use is made of the relation
\begin{equation} \label{27}
\int d^{n}x\, \delta (\sum_{i=1}^{n} x_i)\,
\delta (\sum_{i=1}^{n} x_i^2-R)=
\frac{\pi^{\textstyle\frac{n-1}{2}}}
{\sqrt{n}\,\Gamma(\mbox{\large$\frac{n-1}{2}$})}\,
R^{\textstyle\frac{n-3}{2}} \end{equation}
and the normalization factor is dropped as inessential. It follows from
eqs.~\ref{26} and \ref{16} that in this model
\begin{equation} \label{28}   q(u,t)\sim
t^{n-2}e^{\textstyle{-\frac{n}{2}\,u^2-\frac{n-1}{2}\,t^2}}.
\end{equation}
Inserting this into eq.~(\ref{21}), we obtain
\begin{equation} \label{29}
\rho(\xi)\sim\left(1+\frac{n\xi^2}{n^2-1}\right)^
{\textstyle{-\frac{n}{2}}}, \end{equation}
which is nearly the same as eq.~(\ref{13}). (The exact eq.~(\ref{13})
with $c=1$ and $n$ replaced by $n-1$ would result if the $u$-dependence
in eq.~(\ref{24}) were taken to be $\delta(u)$ ).

Eq.~(\ref{29}) presents a typical appearence of the departure from
normal distribution owing to error uncertainty induced by statistical
reasons only, i.e. by insufficiently large number $n$ of measurements.
Obviously, at $n\to\infty$ the distribution becomes Gaussian, except for
very large $\xi\,\widetilde{>}\,n^{1/4}$ where the correction $\sim\xi^4/n$
becomes essential; at still larger $\xi\,\widetilde{>}\,\sqrt{n}$ the
quadratically-exponential fall-off is in fact changed by a power. In
this model systematic errors are disregarded or, maybe one can say,
simulated by something looking like statistical errors, which does not
seem to be correct, and that probably manifests itself in the necessity
of introducing a scaling parameter~$c$ into eq.~(\ref{13}).

The implication of the considered examples is that no definite
conclusions regarding the probability distribution $w(\xi)$ can be drawn
from eq.~(\ref{21}) without one or other assumption about the function
$q(u,t)$ which embodies information on the mechanism of generating the
error uncertainty. However, there is yet another simple model worth
discussing, based on the hypothesis that the primary source of the
departure from normal distribution hides in the overlooked systematic
errors. It is characterized by the function $q(u,t)$ of the form
\begin{equation} \label{30}
q(u,t)=(\delta(u)+\mu(u))\,\delta(t-1), \end{equation}
where $\mu(u)$ is some smooth function (an overall normalization
coefficient is omitted). Contrary to both models of \cite{2} and
\cite{3}, here the the device dispersion $\sigma$ (the spread of the
distribution~(ref{15}) ) appears to be measured with a good accuracy
($\sigma'=\sigma$ due to the second $\delta$-function and
eq.~(\ref{19}) ), while systematic errors show up in some indefinite
shift $s$ of the measurement scale and are reflected by the addition of
$\mu(u)$ to $\delta(u)$in eq.~(\ref{30}). The additive form of the first
multiplier corresponds to the assumption that all measurements can be
divided into two separate groups of a comparable weight --- the `good'
measurements which are free from systematic errors, and the `bad' ones
which are subject to them, these two groups being represented by two
terms $\delta(u)$ and $\mu(u)$ respectively. Since the systematic errors
are randomly distributed in their values without any visible correlation
to a measured error (which they may exceed many times), the function
$\mu(u)$ can be considered as nearly constant at $u\!\sim\!1$, with some
decrease at very large $u$ rapid enough to provide the convergence of
the normalization integral. So, the dependence on $u$ in eq.~(\ref{30})
looks as a sharp peak at $u\!=\!0$ rising above a smooth, almost
uniformly smeared background. Clearly, the weight of the background is
expected to vary with changing the criteria of the data selection, and
in this sense the above-discussed universality of the distribution
$w(\xi)$ cannot exactly hold; the function $\mu(u)$ can be viewed to
contain an uncertain numerical factor depending on the selection
procedure.

Inserting eq.~(\ref{30}) into (\ref{21}), one can perform the
integration if the function $\mu(u)$ is slowly varying ($\mu'(u)\ll1$),
and get
\begin{equation} \label{31}
\rho (\xi)\approx\frac{1}{\sqrt{2\pi}}\,e^{\textstyle-\frac{\xi^2}{2}}
+\mu(-\xi), \end{equation}
since the Gaussian exponential of eq.~(\ref{21}) acts as a
$\delta$-function when integrated with $\mu(u)$.

The obtained formula is qualitatively similar to the above
eq.~(\ref{10}) --- it represents a sum of two functions, one rapidly and
the other slowly decreasing, and its plot must have a bend near the
value of $\xi$ at which the two terms become equal. One of the pronounced
features of experimental distribution (curve 1 in fig.~2) thus finds
rather simple and natural explanation in the influence of unrevealed
systematic errors. There is however a remarkable difference in the
first term between eqs.~(\ref{31}) and (\ref{10}). In the model just
considered the distribution is distorted only at large $\xi$ and
remains nearly Gaussian to the left of the bending point, while the
experimental curve 1 runs essentially above the Gaussian curve 2 even
at $\xi<3$. Moreover, had the first term of eq.~(\ref{10})
Gaussian-like form, the right-hand nearly horizontal portion of the
curve (at $\xi>3$) would have to go much lower than it really does,
since from eq.~(\ref{3}) $w(3)\approx0.003$ instead of actual 0.05.
This disparity may look not serious, as it can be easily eliminated by
an appropriate smearing of the $\delta$-functions in eq.~(\ref{30}).
However, such way of doing is not quite satisfactory because of its
arbitrariness and indeterminacy, it leaves too much freedom in the
choice of the function $q(u,t)$ and does not allow to say anything
definite about what the distribution must be like for theoretical
reasons. Nor is it of any use in clarifying the origin of the unit
coefficient in the exponent of eqs.~(\ref{10}) and (\ref{4}) which is a
pure accident in this scheme. Therefore it makes sense to look for some
other sources of possible distribution deformations which could be
effective at $\xi<3$ and maybe shed light on the latter problem.

Eq. (\ref{21}) can be easily extended to the case of the probability
density $p(x,a)$ in eq.(\ref{15}) having a more general than Gaussian
form  \begin{equation} \label{32}
p(x,a)=\frac{1}{\sigma}\,p\left(\frac{x-a-s}{\sigma}\right),
\end{equation}
then \begin{equation} \label{33}
\rho(\xi)=\int du\,dt\,q(u,t)\,tp(u+t\xi).\end{equation}

If the function $p(y)$ is rapidly decreasing at large argument, the
hypothesis of eq.~(\ref{30}) yields a similar generalization of
eq.~(\ref{31}):
\begin{equation} \label{34}
\rho(\xi)\approx p(\xi)+\mu(-\xi). \end{equation}
Again we see that at $\xi$ not very large where the first term is
dominant (the boundary value of $\xi$ depending on the selection
criteria), the experimentally measured distribution duplicates the
primary scatter caused by the device, and we come back to the problem
of explaining its non-Gaussian character. So, in what follows we ignore
the mesurement uncertainty of the function $p(\xi)$, taking $q(u,t)$ to be
defined by eq.~(\ref{22}), and concentrate on examining a possible form
of $p(\xi)$ itself, which is then the same as $\rho(\xi)$.

\section {The central limit theorem}

Normal distribution, be it justified or not, holds a specific position
in statistics and is often favoured over other distributions in default
of information. So, clearing up the question why should (or shouldn't)
some probability distribution be normal, it is worthwhile to recall the
central limit theorem which is in fact the only argument for that. A
simple version of the central limit theorem (there are many of them,
see e.g. \cite{1}) can be formulated as follows.

Let $x_1, x_2, ... x_n$ be $n$ independent random quantities with
distribution density of $x_k$ being $\rho_k(x_k)$. Let three lowest
moments of each distribution $\rho_k$ exist:
\begin{equation} \label{35} \mbox{\bf M}\,x_k=\overline{x}_k,\;\;\;
\mbox{\bf M}\,(x_k-\overline{x}_k)^2=\sigma_k^2,\;\;\;
\mbox{\bf M}\,|x_k-\overline{x}_k|^3=t_k^3 \end{equation}
(here $\mbox{\bf M}f$ denotes mean value of $f$). Construct a new
variable $y=\sum_{k=1}^n x_k$ having the distribution density
\begin{equation} \label{36}
\rho(y)=\int dx_1...dx_n\,\rho(x_1)...\rho(x_n)\,
\delta(y-\sum_{k=1}^n x_k) \end{equation} with the moments
\begin{equation} \label{37}
\begin{array}{c}\mbox{\bf M}\,y=
\displaystyle\sum_{k=1}^n\overline{x}_k=\overline{y},\;\;\;
\mbox{\bf M}\,(y-\overline{y})^2=\sum_{k=1}^n\sigma_k^2=\sigma^2,\\
\mbox{\bf M}\displaystyle\sum_{k=1}^n|x_k-\overline{x}_k|^3=
\sum_{k=1}^n t_k^3=t^3, \end{array}\end{equation}
and consider the limit $n\to\infty$. The theorem states that if the
requirement
\begin{equation} \label{38}
\lim_{n\to\infty}\frac{t}{\sigma}=0 \end{equation}
(Liapunov's condition) is satisfied, then, whatever the distributions
$\rho_k (x_k)$ are, the distribution $p(y)$ tends to Gaussian with mean
value $\overline{y}$ and dispersion~$\sigma$:
\begin{equation} \label{39}
\lim_{n\to\infty}p(y)=\frac{1}{\sqrt{2\pi}\,\sigma}\,
e^{\mbox{\large $-\frac{(y-\overline{y})^2}{2\sigma^2}$}}.
\end{equation}

The meaning of the condition (\ref{38}), roughly speaking, is that it
necessitates fluctuations of all $x_k$ to make comparable contributions
into fluctuations of~$y$, preventing from only a small number of them
being really effective. In a typical case with all $\sigma\!_k$ of the
same order, $\sigma\sim\sigma\!_1\,n^{1/2},\; t\sim\sigma\!_1\,n^{1/3},
\linebreak t/\sigma\sim~n^{-1/6}\to 0$. (The overall factor
$\sigma\!_1$ itself is supposed to be $\sim n^{-1/2}$, so that $\sigma$
tends to a constant at $n\to\infty$).

For logical coherency we sketch here an outline of the proof.
Introducing Fourier transforms
\begin{eqnarray}
\int\!dx\,e^{\mbox{\small${iux}$}}\rho_k(x)=\phi_k(u), \\ \label{40}
\int\!dy\,e^{\mbox{\small${iuy}$}}p(y)=\Phi(u), \label{41}
\end{eqnarray}
we have from eq. (\ref{34})
\begin{equation} \label{42}
\Phi(u)=\prod_{k=1}^n \phi_k (u),\;\;\;\;
\ln\Phi(u)=\sum_{k=1}^n \ln\phi_k(u).\end{equation}
With eq.~(\ref{38}) being fulfilled, each $\rho_k$ at $n\to\infty$ is
shrinking towards its maximum at $x=\overline{x}_k$ and the exponential
in the integrand of eq.~(\ref{40}) can be expanded near that point to
yield \begin{eqnarray}
\phi_k(u)\approx e^{\mbox{\small${iu\overline{x}\!_k}$}}\,
(1-\mbox{\large{$\frac{u^2}{2}$}}\,\sigma\!_k^2)\;,\\ \label{43}
\ln\phi_k(u)\approx iu\overline{x}\!_k
-\mbox{\large$\frac{u^2}{2}$}\,\sigma\!_k^2\ . \label{44}
\end{eqnarray}
Inserting eq. (\ref{44}) into (\ref{42}) and using eqs.~(\ref{37}), we
obtain
\begin{equation} \label{45}
\ln\Phi(u)\approx iu\overline{y}
-\mbox{\large$\frac{u^2}{2}$}\,\sigma^2.
\end{equation}
An important point is that the cubic and all higher terms in the
right-hand side vanish at $n\to\infty$ due to eq.~(\ref{38}).
Exponentiating eq.~(\ref{45}) and making inverse Fourier transform of
$\Phi(u)$ according to eq.~(\ref{41}), we get the Gaussian
expression~(\ref{39}) for $p(y)$.

It may be remarked that the statement of the theorem is true also in a more
general case when $y$ is not an exactly linear function of the variables~ $
x_k $. It is sufficient for this function to be smooth in the vicinity of
the point $x_k=\overline{x}_k$ and exhibit essential changes only with the
departure from this point at distances much greater than the dispersions $
\sigma_k$, so that it could be well approximated by its linear expansion in
differences $(x_k-\overline{x}_k)$ (the coefficients of this expansion can
easily be renormalized to unity by redefinition of $x_k$).

Another possible way of extension of the theorem is associated with
abandoning the condition of independence of random variables~$x_k$. If they
are correlated, then the product $\rho _1(x_1)...\rho _n(x_n)$ in the
integrand of eq. (\ref{36}) which is effectively proportional to
\[e^{\mbox{\scriptsize$\displaystyle -\frac
12\sum_{k=1}^n\frac{(x_k-\overline{x}_k)^2}{\sigma_k^2}$}}\]
is expected to be replaced by
\begin{equation}
e^{\mbox{\small$-\frac 12T_{kl}\left( x_k-\overline{x}_k\right)
\left( x_l-\overline{x} _l\right)$}},  \label{46}
\end{equation}
where $T_{kl}$ is some non-diagonal positive definite constant $n\times n$
matrix (Liapunov's condition (\ref{38}) is supposed to be properly
modified). It is clear that such replacement does not affect the form of the
function $\rho (y)$ which remains Gaussian.

If now a random quantity has a distribution different from Gaussian, it
would be well to conceive which of the conditions of the central limit
theorem is violated. There are not many possibilities for that. First one,
most obvious and suggesting itself, can be that the number $n$ of effective
disturbing factors is not large or only a few of them fluctuate
significantly. A typical example is when all dispersions $\sigma _k$ are
negligible in comparison with one $d_1$ and the sum in the second eq. (\ref
{37}) is dominated by one term $\sigma _1^2$, so that all $x_k$ except $x_1$
can be considered as constants and the distribution of a quantity $y$
coincides in essence with that of $x_1$ up to some shift. If $n$ is large
but finite, then the dropped cubic in $u$ term in eq.~(\ref{45}) is
generally of order $n(\sigma _1u)^3\sim (\sigma u)^3/\sqrt{n}\,$, and since
from eq.~(\ref{41}) (or its inverse)
\begin{equation}
u\sim \frac{y-\overline{y}}{\sigma ^2},  \label{47}
\end{equation}
the Gaussian law is expected to break down at
\begin{equation}
\xi =\frac{y-\overline{y}}\sigma \,\widetilde{>}\,n^{1/6},  \label{48}
\end{equation}
where this term becomes of order or more than 1. (This condition is replaced
by $\xi \,\widetilde{>}\,n^{1/4}$ with the analogous estimate of a quartic
instead of cubic term in the case when the latter happens to vanish). It is
seen that the number $n$ must be extremely large to provide the validity of
the standard law even only for $\xi \leq 4$.

Another natural possibility for a quantity $y$ to have non-Gaussian
distribution consists in the nonlinear dependence of $y$ on the disturbing
factors $x_k$. The usual way of reasoning is that if the number $n$ of
$x_k$'s is large and each partial dispersion $\sigma _k$ is small $\sim
\sigma n^{-1/2}$, then any smooth function $y(\{x_k\})$ varying
at scales of $x_k\sim \sigma $ can be linearized for small
variations of $x_k\sim \sigma _k $. However, it is not true,
which is seen from the fact that if $y$ is normally distributed
then $z=f(y)$ is generally not. This paradox emerges from the
invalidity of a linear approximation for the function
$y(\{x_k\})$ when it changes essentially only in a small number
of directions in the multidimensional space of $\{x_k\}$; in this
case largeness of $n$ is not a sufficient condition for the
nonlinear terms to be negligible, in fact their total
contribution is comparable in value with the linear part of the
expansion. So, the nonlinearity may matter when the measured quantities are
being chosen at random, without any consideration of their dependence on the
disturbing factors. We shall not discuss this point in more detail since one
is not led here to any definite conclusion about the character of influence
of this effect on the resulting distribution; it can only be thought to
weaken with averaging over a variety of dissimilar data, while in
distributions of measurements of one or a few analogous quantities it may
appear more important. (This is yet another possible reason for violation of
the universality discussed in Sec. 2).

There is a further way worth considering to get a distribution different
from Gaussian, which will be given more attention in the next section; it is
associated with the possibility that the variables $x_k$ might be correlated
without representation (\ref{46}) for their combined probability to hold. It
can be realized e.g. by supposing the matrix $T$ in~(\ref{46}) to
depend on the sum of $x_k$'s which can be not small in spite of
smallness of $x_k$'s owing to large $n$. A reason for such
assumption is provided by observation that the accuracy of a
measuring device is mostly not constant but different in various
parts of the scale, usually best in the middle and worsening to
the extremities. This approach seems to have more definite
consequences as to the resulting data distribution than the
previous ones, which will be seen from a model considered below.

\section{The variable dispersion effect}

Let the total set of variables $x_k$ determining the measurement
result $y$ consist of two subsets of $n$ and $n_1$ variables
respectively, the former $n$ being entirely independent and the
latter $n_1$ entangled together. Both numbers $n$ and $n_1$ are
supposed to be large enough so that all $\rho_k(x_k)$ could be taken
in the Gaussian form and thus
\begin{eqnarray} \label{49}
\rho(y)=(2\pi)^{\textstyle{-\frac{n+n_1}{2}}}\displaystyle\int d^n x
\left(\,\prod_{k=1}^n\frac 1{\sigma_k}
e^{\textstyle-\frac{x_k^2}{2\sigma_k^2}}\right)\cdot\nonumber\\
\displaystyle\int d^{n_1}x\,(\det T)^{1/2}\,e^{\textstyle-\frac12
T_{kl}x_k x_l}\,\delta\!\left(y-\displaystyle\sum_{k=1}^{n+n_1}x_k\right),
\end{eqnarray}
where $T_{kl}x_k x_l$ implies summation over $k,l$ running from $n+1$ to
$n+n_1$. The mean values $\overline{x}_k$ are put to be zero by making
proper shifts of $x_k$'s.

Taking the inner integral over $d^{n_1}x$, we obtain
\begin{eqnarray}
&\rho(y)=(2\pi)^{\textstyle-\frac n2}\displaystyle\int d^n x
\left(\prod_{k=1}^n\frac 1{\sigma_k}
e^{\textstyle-\frac{x_k^2}{2\sigma_k^2}}\right)\,
g\!\left(y-\displaystyle\sum_{k=1}^{n}x_k\right),\\ \label{50}
&g(z)=\displaystyle \frac 1{\sqrt{2\pi}\,\sigma_0}\,
e^{\textstyle-\frac{x_0^2}{2\sigma_0^2}},\\ \label{51}
&\sigma_0^2=\displaystyle\sum_{kl}(T^{-1})_{kl}. \label{52}
\end{eqnarray}

Eq. (\ref{50}) is very similar to eq.~(\ref{36}) with Gaussian-like
probabilities~$\rho_k(x_k)$; an important difference is that
$\delta$-function in the integrand has been changed by Gaussian-like
function $g(z)$, eq.~(\ref{51}). This expression looks as a natural
generalization of eq.~(\ref{36}) for $\rho(y)$ when $y$ is not
exactly equal to the sum of $x_k$ but is a random function whose
values are normally distributed about this sum with the
dispersion~$\sigma_0$. Eq.~(\ref{50}) may be postulated as a
hypothesis itself, without reference to eq.~(\ref{49}); it seems
possible for it to have more general character and to be produced by
some other underlying mechanisms.

It is seen that the form of the matrix $T_{kl}$ in eq. (\ref{49}) does not
matter much, it might well be diagonal. A significant point in what follows
is the assumption that this matrix and hence the width $\sigma _0$ of the
peak of the function $g(z)$ may depend on $(\sum_{k=1}^nx_k)$, so
that further integration in eq.~(\ref{50}) becomes nontrivial.
This dependence is supposed to model the above-mentioned
variation of the measurement dispersion within the scale of a
device. It should be emphasized that the two groups of variables
$x_k$ ($k=1$ to $n$ and $k=n+1$ to $n+n_1$) play different
parts in this model even in the case of a diagonal matrix
$T_{kl}$ which is taken to depend only on the first group of
$x_k$'s. A possible way of interpreting the second group of $n_1$
variables is to consider them as the remnants of the corrected
systematic errors. The correction that an experimenter has to
introduce when he detects some systematic effect, while supposed to
be a constant, may in reality depend on the measured value, and this
fact can be simulated by the specified variability of $\sigma_0$.

Now, to calculate $\rho (y)$, let us make again a Fourier transform of
eq.~(\ref{50}), as we did it earlier with eq.~(\ref{34}):
\begin{eqnarray}
\Phi (u) &=&\int dy\,e^{iuy}\rho (y)=(2\pi )^{\textstyle -\frac n2}\int
d^nx\,\left( \prod_{k=1}^n\frac 1{\sigma _k}\right) \cdot   \nonumber \\
&&\int dh\,\delta\!\left( h-\sum_{k=1}^nx_k\right) \,e^{
\mbox{\footnotesize$\displaystyle
-\frac 12\sum_{k=1}^n\frac{x_k^2}
{\sigma _k^2}+iuh-\frac{u^2}2\sigma _0^2(h)$}},  \label{53}
\end{eqnarray}
where a new integration variable $h=\sum_{k=1}^n$ is introduced.
Substituting for $\delta $-function its Fourier representation, we can
integrate over $x_k$ and get
\begin{eqnarray}
\Phi (u) &=&\frac 1{\sqrt{2\pi}\,\sigma }\int dh\,e^{
\mbox{\scriptsize$\displaystyle
-\frac{h^2}{2\sigma^2}+\mbox{\footnotesize$iuh$}-
\frac{u^2}2\sigma _0^2(h)$}},  \label{54} \\
\sigma ^2 &=&\sum_{k=1}^n\sigma _k^2.  \label{55}
\end{eqnarray}
Performing now inverse Fourier transform of eq. (\ref{54}), we obtain
\begin{equation}
\rho (y)=\frac 1{2\pi \sigma }\int \frac{dh}{\sigma _0(h)}\,e^{
\mbox{\scriptsize$\displaystyle
-\frac{h^2}{2\sigma^2}-\frac{(h-y)^2}{2\sigma _0^2(h)}$}}.  \label{56}
\end{equation}

In the case of $\sigma _0(h)=const$  this formula obviously gives the
Gaussian expression for $\rho (y)$ with total dispersion $\Delta =\sqrt
{\sigma ^2+\sigma _0^2}$. If $\sigma _0(h)$ is a slowly varying
function, then the integral is nearly the same but with $\Delta $
depending on $y$ via $\sigma _0(h)$, where $h$ is a function of
$y$ determined from the saddle point equation
\begin{equation}
h=y\,\frac{\sigma ^2}{\sigma ^2+\sigma _0^2}\:.  \label{57}
\end{equation}
Typically $\sigma _0(h)$ is a growing function (at $h>0$), and since
$h$ is nearly proportional to $y$ from eq.~(\ref{57}), $\Delta $ also
grows with $y$, which means that the distribution fall-off is slower
than by Gaussian law.

It makes sense to consider more closely a case of a quadratic
function~$\sigma_0^2(h)$:
\begin{equation} \label{58}
\sigma_0^2(h)=\sigma_0^2+\gamma(h-h_0)^2,\;\;\; \gamma>0 \end{equation}
($\gamma,\sigma_0,h_0$ are some constants), which is in fact the
simplest suitable form, since a linear function is not appropriate
because of its sign reversal at some point. Eq. (\ref{58}) can be
viewed not merely as a Taylor expansion of some smooth function at
small deviations $(h-h_0)$, but more generally as a reasonable
approximation for any function with a similar behaviour in a bounded
region, i.e. having a minimum and growing progressively in both
directions from this minimum.

An important property of the quadratic function $\sigma_0^2(h)$ is
that it yields the simple-exponential fall-off of the distribution
density $\rho(y)$ at asymptotically large~$y$. Indeed, the exponent
in the integrand of eq.~(\ref{56}) with regard to eq.~(\ref{58}) at
$y\to\infty,\,h\to\infty$ and $h\ll y$ (this will be seen to hold for
the saddle point) turns into
$$-\frac{h^2}{2\sigma^2}-\frac{y^2}{2\gamma h^2}\;.$$
This function of $h$ has a sharp maximum at
$h=\gamma^{-1/4}\sqrt{\sigma y}$ and its value at this point is
$\;-y/(\sigma\sqrt\gamma)$, so
\begin{equation} \label{59}
\rho(y)\sim e^{\mbox{\large$-\frac{y}{\sigma\sqrt\gamma}$}}
\;\;\; (y\to\infty) \end{equation}
with main exponential accuracy (without pre-exponential power factors).

Another interesting consequence of the assumption~(\ref{58}) for
$\sigma_0^2(h)$ is the possibility to calculate explicitly from
eq.~(\ref{54}) the Fourier transform of the distribution density:
\begin{equation} \label{60}
\Phi(u)=\frac 1{\sqrt{1+\gamma\sigma^2 u^2}}\:
e^{\textstyle-\frac{u^2}2\left(\sigma_0^2+\gamma h_0^2+
\sigma^2\mbox{\large$\frac{(1-i\gamma h_0 u)^2}
{1+\gamma\sigma^2 u^2}$}\right)} \end{equation}
and hence the moments $\int\!\rho(y)y^m dy$, which are in fact the
coefficients of the Taylor expansion of $\Phi(u)$. In particular, for
the dispersion $\Delta$ we have
\begin{equation} \label{61}
\Delta^2=\int\!\rho(y)y^2dy=-\Phi''(0)=
\sigma_0^2+\gamma h_0^2+\sigma^2(1+\gamma)\,.
\end{equation}

It should be noted that eq.~(\ref{56}) is not yet just the expression
to be compared with the experimental data distributions in figs.~1--3,
since it is supposed to refer to a certain device or setup, while
experimental curves involve the results obtained on a large number of
them. To get from eq.~(\ref{56}) the measured distribution, we must
average it over the entering parameters $\sigma, \sigma_0, h_0,
\gamma$ peculiar to any given device, with some weight $P$
characterizing the relative probability of finding in the world a
device (setup) with certain values of the parameters. Moreover,
suppose we are dealing (as was said above) only with `good'
measurements, whose reported errors reflect quite adequately the
dispersion of the setup. Then we can obtain the distribution density
$\rho(y,\Delta)$ for the data with a given error $\Delta$ by
integrating over a range of the parameters subject to
constraint~(\ref{61}). Introducing slightly redefined parameters
\begin{equation} \label{62}
b=h_0\sqrt{\gamma},\;\;\;c=\sigma\sqrt{\gamma}\end{equation}
in place of $h_0$ and $\gamma$, we can write by eqs.~({\ref{56}) and
(\ref{58})
\begin{eqnarray} \label{63}
\rho(y,\Delta)=\int dS_3\,P(\sigma,\sigma_0,b,c)\,\frac1{2\pi\sigma}
\int\frac{dh}{\sqrt{\mbox{\footnotesize$\sigma_0^2+
(\displaystyle{\frac{ch}{\sigma}}-b)^2$}}}\,\cdot\nonumber\\
\exp\mbox{\large${\left[-\frac{h^2}{2\sigma^2}-\frac{(y-h)^2}
{2(\sigma_0^2\,+({\textstyle\frac{ch}{\sigma}}
-\,b)^2)}\right]}$}\,,   \end{eqnarray}
where the outer integral is to be taken over three-dimensional
surface of a hypersphere
\begin{equation} \label{64}
\sigma^2+\sigma_0^2+b^2+c^2=\Delta^2  \end{equation}
in the space of the parameters $(\sigma,\,\sigma_0,\,b,\,c)$ (more
accurately, the integration region covers 1/8 part of the hypersphere
where $\sigma>0,\sigma_0>0, c>0$). To put it differently,
\begin{equation} \label{65}
\int dS_3=\int d\sigma\,d\sigma_0\,db\,dc\,
\delta(\Delta-\sqrt{\sigma^2+\sigma_0^2+b^2+c^2})\,.
\end{equation}
The distribution of devices $P(\sigma,\sigma_0,b,c)$ in
eq.~(\ref{63}) is supposed to obey the normalization condition
\begin{equation} \label{66}
\int P\,d\sigma\,d\sigma_0\,db\,dc=\int P\,d\Delta\,dS_3=1.
\end{equation}

The experimentally studied quantity is not quite $\rho(y,\Delta)$ but
rather the distribution in $\xi=y/\Delta$:
\begin{equation} \label{67}
\rho(\xi)=\int dy\,d\Delta\,\rho(y,\Delta)\,\delta(\xi-\frac y\Delta)
=\int\Delta\,d\Delta\,\rho(\xi\Delta,\Delta)\,.
\end{equation}
Inserting here eq.~(\ref{63}) and scaling out the variable~$\Delta$,
\begin{equation} \label{68}
\begin{array}{c}
\sigma=z_1\Delta,\;\sigma_0=z_2\Delta,\;b=z_3\Delta,\;c=z_4\Delta,\\
h=t\sigma=tz_1\Delta,\;dS_3=\Delta^3d\Omega,\end{array}
\end{equation}
we obtain
\begin{eqnarray}
&\rho(\xi)={\displaystyle\int}d\Omega\,\overline{P}(\mbox{\bf z})
\rho_0(\xi,\mbox{\bf z}), \label{69}\\
&\rho_0(\xi,\mbox{\bf z})=\displaystyle\frac 1{2\pi}\int\frac{dt}
{\sqrt{z_2^2+(tz_4-z_3)^2}}\,e^{\mbox{\large$-\frac{t^2}2-
\frac{(\xi-tz_1)^2}{2[z_2^2+(tz_4-z_3)^2]}$}}, \label{70}\\
&\overline{P}(\mbox{\bf z})={\displaystyle\int}d\Delta\,\Delta^3
P(z_1\Delta,z_2\Delta,z_3\Delta,z_4\Delta),\;\;\;
{\displaystyle\int}d\Omega\,\overline{P}(\mbox{\bf z})=1,\label{71}
\end{eqnarray}
where now \mbox{\bf z} is a unit four-dimensional vector and
$d\Omega$ an element of the surface on a unit hypersphere,
$\sum_{i=1}^4 z_i^2=1$. The $d\Omega$ integration in
eqs.~(\ref{69}) and (\ref{71}) goes over the region
$-1\le z_3\le 1,\; 0\le z_{1,2,4}\le 1$.

Eq. (\ref{69}) seems to be of little use without knowing the
distribution of devices $P(\sigma,\sigma_0,b,c)$ which is rather
indefinite, just like it was with eq.~(\ref{21}) containing an
unknown function $q(u,t)$. However, there turns out to be a
substantial difference between these two formulae: whereas
eq.~(\ref{21}) is sensitive to the form of the function $q(u,t)$ and
the details of its behaviour, it is quite another matter with the
function $\overline{P}(\mbox{\bf z})$ in eq.~(\ref{69}) where the
other multiplier $\rho_0(\xi,\mbox{\bf z})$ in the integrand is a
function of $\mbox{\bf z}$ sharply peaked at the spherical pole
$z_4=1, z_1=z_2=z_3=0$.

To be more precise, consider the function $\rho_0(\xi,\mbox{\bf z})$
for $\xi\gg 1$. Unless $z_4\to~0$, the integral in eq.~(\ref{70})
then has two saddle points close to $t=\pm\sqrt{\xi/z_4}$ and, in
perfect analogy with eq.~(\ref{59}),
\begin{equation} \label{72}
\rho_0(\xi,\mbox{\bf z})\sim e^{\mbox{\large$-\frac{\xi}{z_4}$}}\,.
\end{equation}
(The case of $z_4\to 0$ can be disregarded, since the integral then
becomes Gaussian and $\rho_0(\xi,\mbox{\bf z})$ falls off rapidly as
$\exp(-\xi^2/2)$ giving a negligible contribution into the $d\Omega$
integral of eq.~(\ref{69}) ). We conclude that $\rho_0(\xi,\mbox{\bf z})$
at large $\xi$ has a sharp maximum at $z_4=1$ and decreases rapidly
with the departure from this pole. It means that that if we suppose
$\overline{P}(\mbox{\bf z})$ to be more or less smooth on the sphere,
the integral in eq.~(\ref{67}) will be dominated by a small region
around this pole where $\overline{P}(\mbox{\bf z})$ can be replaced
by a constant, so that the details of the distribution
$\overline{P}(\mbox{\bf z})$ become inessential, and we get
\begin{equation} \label{73}
\rho(\xi)\sim e^{-\,\xi}. \end{equation}
This is the result striven for: not only the distribution density
$\rho(\xi)$ decays exponentially, but even with the unit coefficient
in the exponent, just as it was experimentally found. True, the
justification holds only for $\xi\gg 1$ and depends on some
hypothesis as to the form of the distribution of devices
$P(\sigma,\sigma_0,b,c)$; but this hypothesis does not seem very
restrictive, and the asymptotic form (\ref{73}) of $\rho(\xi)$ may be
hoped to keep approximate validity up to $\xi\sim 1$ as is often the
case.

It is worthwhile to refine eq.~(\ref{73}) with the inclusion of a
pre-exponential (power) factor. With this aim in view, let us rewrite
eq.~(\ref{68}) in the large~$\xi$ limit with more accuracy. It follows
from eq.~(\ref{72}) that $z_4$ can be assumed to differ from 1 only
by a quantity of order $1/\xi$ or less, for beyond that region
$\rho_0(\xi,\mbox{\bf z})$ vanishes more rapidly than
$\sim\exp(-\xi)$ and can be put equal to zero when integrated over
$d\Omega$ in eq.~(\ref{67}). So we shall take
\begin{equation} \label{74}
1-z_4\approx\frac 12(z_1^2+z_2^2+z_3^2)\,\widetilde{<}\,\frac 1\xi\;,
\end{equation}
which means that $z_1,z_2,z_3\,\widetilde{<}\,\xi^{-1/2}$.
Putting $$t=\pm\sqrt{\frac{\xi}{z_4}}+\tau $$ and expanding the
exponent in the integrand of eq.~(\ref{70}) in $\tau$, one can check
that the quadratic term is of order $\tau^2/\xi$ and so
$\tau\sim\xi^{-1/2}$. Thus, keeping the terms $\sim 1$, we can write
eq.~(\ref{70}) in an approximate form:
\begin{eqnarray}
\rho_0(\xi,\mbox{\bf z})\approx\displaystyle\frac 1{2\pi}\int
\frac{dt}{|t|}\,e^{\mbox{\large$-\frac{t^2}2-\frac{\xi^2}{2t^2z_4^2}\,
 (1-\frac{2tz_1}{\xi}+\frac{2z_3}{tz_4})$}} \nonumber\\
\approx\frac 1{\sqrt{2\pi\xi}}\; e^{\mbox{\large$-\frac{\xi}{z_4}$}}\;
\cosh(\sqrt\xi(z_1-z_3))\,. \label{75} \end{eqnarray}
A cosh function arises as a result of adding together the
contributions of the two saddle points, $t=\pm\sqrt{\xi/z_4}$.

Inserting this expression into eq.~(\ref{70}) and supposing
$\overline{P}(\mbox{\bf z})$ to be smooth and nonvanishing at the
point $z_4=1$, we get
\begin{equation} \label{76}
\rho(\xi)\approx\frac
1{\sqrt{2\pi\xi}}\,e^{-\,\xi}\,\overline{P}(0,0,0,1)
\int\!d\Omega\,e^{{\textstyle-\frac{\xi}2}(z_1^2+z_2^2+z_3^2)}
\cosh(\sqrt\xi(z_1-z_3))\,.\end{equation}
For small $z_1,z_2,z_3\sim\xi^{-1/2}$, the sphere can be approximated
by a plane and thus $d\Omega$ replaced by $dz_1\,dz_2\,dz_3$, so
the integral in eq.~(\ref{76}) is proportional to $\xi^{-3/2}$ and
finally
\begin{equation} \label{77}
\rho(\xi)\sim\frac 1{\xi^2}\,e^{-\,\xi}\;\;(\xi\to\infty)\,.
\end{equation}
The omitted multiplicative constant includes the uncertain
quantity\linebreak $\overline{P}(0,0,0,1)$ with a definite numerical
factor, easily computable but not too interesting.

It must be admitted that the assumption of $\overline{P}(\mbox{\bf
z})\to const$ at $z_4\to1$ which is of importance here, is taken
rather arbitrarily and it is difficult to say what the behaviour of
the function $\overline{P}(\mbox{\bf z})$ at $z_4\to 1$ should be
like in reality. If, in particular, $\overline{P}(\mbox{\bf
z})\sim(1-z_4)^\nu$ or, more generally, is a homogeneous function of
the power $2\nu$ in $z_1,z_2,z_3$, then $\xi^{-2}$ in
eq.~(\ref{77}) is changed by $\xi^{-2-\nu}$. (In terms of the
parameters entering eq.~(\ref{58}) this implies the distribution
behaving powerwise with $\gamma$ at $\gamma\to\infty$). The main
point however is that the exponential $\exp(-\xi)$ is unaffected by
these variations even in the case of a more rapid than power-like
vanishing of $\overline{P}(\mbox{\bf z})$ at $z_4=1$.

Of some interest may be the limiting case when $\overline{P}(\mbox{\bf z})$
is concentrated near the point $z_4=1$, i.e.
\begin{equation} \label{78}
\overline{P}(\mbox{\bf z})=\delta(z_1)\,\delta(z_2)\,\delta(z_3),
\end{equation}
which corresponds to the situation of all devices having very large
$\gamma$ in eq.~(\ref{58}) and $\xi\ll\gamma$. Then eqs.~(\ref{69})
and (\ref{70}) give
\begin{equation} \label{79}
\rho(\xi)=\frac 1\pi\int_0^\infty\frac{dt}t\,
e^{\mbox{\normalsize$-\frac{t^2}2-\frac{\xi^2}{2t^2}$}}=
\frac 1\pi K_0(\xi), \end{equation}
where $K_0$ is the McDonald function (modified Bessel function of the
second kind). This expression holds for $\xi\sim1$; it fails at
$\xi\widetilde{>}\gamma$ because then the range of variation of
$z_{1,2,3}\sim\xi^{-1/2}$ becomes comparable with (or less than) the
smearing of $\delta$-functions in eq.~(\ref{78}), which by
eqs.~(\ref{68}) and (\ref{61}) has the order
$\sigma/\Delta\sim\gamma^{-1/2}$. Asymptotic form of eq.~(\ref{79})
at $\xi\gg 1$
\begin{equation} \label{80}
\rho(\xi)\approx\frac 1{\sqrt{2\pi\xi}}\,e^{-\,\xi}
\end{equation}
is similar to eq.~(\ref{77}) but naturally differs in pre-exponential
power of $\xi$.

Eq. (79) breaks down also at $\xi\to 0$ where $K_0(\xi)$ has a
logarithmic singularity; it ceases to hold when
$\xi\sim\gamma^{-1/2}$ so that again the smearing of
$\delta$-functions becomes effective (we suppose $z_1\sim z_2\sim
z_3\sim\gamma^{-1/2}$, i.e. $\sigma\sim\sigma_0\sim h_0\sqrt\gamma$).
A real distribution density is finite at $\xi=0,\; \rho(0)\sim
\ln 1/\gamma$. Thus the range of validity of eq.~(\ref{79}) is
\begin{equation} \label{81}
\frac 1{\sqrt\gamma}\ll\xi\ll\gamma. \end{equation}
Here $\gamma$ represents the minimal value of this parameter among
all devices, which is supposed to be $\gg 1$; this case does not look
realistic but is considered to illustrate the weak dependence of
$\rho(\xi)$ on the distribution of devices.

\section{Conclusion}

Main results of this work are somewhat different from those of the
previous work \cite{2} and can now be summed up in brief as follows.

First, estimates of probability of deviation of the measured from
true value of some quantity at large $\xi\;(\widetilde{>}4)$ do not
seem to make any sense at all without supplementary information,
neither by Gaussian nor by any other similar formulae. There is no
universality at large $\xi$; typical is the existence of a critical
value of $\xi$ (dependent on the sample of data) beyond which the
deviation probability almost ceases to fall down. This slowly
dropping large~$\xi$ tail of a distribution is natural to associate
with the contribution of the measurement results perverted by the
overlooked systematic errors (`bad' measurements).

Second, there is a conspicuous departure from the standard law also
at smaller $\xi$, having a different form and hardly attributable to
the same effect which is expected to be negligible at small $\xi$.
Among other explanations, the effect of the error uncertainty
resorted to in earlier papers \cite{2,3} does not seem to be an apt
one, for its manifestations are too indefinite and unspecific, while
the experimental distribution in this region shows an intriguing
correspondence with the simple exponential $\sim\exp(-\xi)$.

Third, a new effect has been found (the variable dispersion effect)
capable of inducing the distribution to have asymptotic behaviour
$\rho(\xi)\sim\exp(-\xi)$ at large $\xi$, with some less definite
pre-exponential (power) factor depending on the distribution of
measuring devices over their parameters. The effect takes into
account a possibility of variation of the device dispersion with
changing value of the measured quantity and may occur as a
manifestation of the detected and corrected systematic errors, even
in the cases of `good' measurements.

It should be noted that the approach applied here is merely a model,
by no means the only possible. There can exist several other sources
of deviation from the standard law---the error uncertainty or those
mentioned in Sec. 4. In particular, the former is hardly believable
to have no appreciable influence, since the experimenters usually do
not know their measurement errors with much accuracy. It may happen
that various effects compensate each other and leave alone the
variable dispersion effect, but one cannot say for sure if it is
really so. Maybe the question could be elucidated with more abundant
statistics and more detailed study of distributions in different
subsamples. It seems of interest also to study the analogous
distributions of data obtained in other areas of science, not only in
elementary particle physics and not only in phisics at all. The
questions touched on here are of evident practical importance when
any estimates of probability of some events are involved; one should
have a clear notion of what the degree of validity of such estimates
is like.
\vspace{5mm}

The author is grateful to Profs. H.Dahmen, S.Brandt and L.Lipatov for
valuable discussions and much assistance in a visit to Siegen
University.

This work was supported partly by the Russian Fund of Fundamental
Researches\linebreak (grant 92-02-16809) and by INTAS (grant 93-1867).

\end{document}